\documentclass{pasj00}
\begin{document}
\SetRunningHead{Y. Takeda and M. Takada-Hidai}{[S/Fe] of Very 
Metal-Poor Stars Revisited}
\Received{2011/09/08}%{yyyy/mm/dd}
\Accepted{2011/11/07}%{yyyy/mm/dd}

\title{Behavior of [S/Fe] in Very Metal-Poor Stars\\
from the S~I 1.046~$\mu$m Lines Revisited
\thanks{Based on data collected at Subaru Telescope, which is operated by 
the National Astronomical Observatory of Japan.}
}

%%% Please use the following style in case that sorting by 
%%% affilation is impossible. 
%
% \author{%
%   D-Firstname \textsc{D-Familyname}\altaffilmark{1}
%   E-Firstname \textsc{E-Familyname}\altaffilmark{1,2}
%   and
%   F-Firstname \textsc{F-Familyname}\altaffilmark{2}}
% \altaffiltext{1}{Address of Institute}
% \email{ddddd@xxx.xxx.xx.xx}
% \email{eeeee@xxx.xxx.xx.xx}
% \altaffiltext{2}{Address of Institute}

\author{Yoichi \textsc{Takeda}}
\affil{National Astronomical Observatory of Japan 
2-21-1 Osawa, Mitaka, Tokyo 181-8588}
\email{takeda.yoichi@nao.ac.jp}
\and 
\author{Masahide \textsc{Takada-Hidai}}
\affil{Liberal Arts Education Center, Tokai University, 
4-1-1 Kitakaname, Hiratsuka, Kanagawa 259-1292}
\email{hidai@apus.rh.u-tokai.ac.jp}

%%% Please use the following style in case that sorting by 
%%% affilation is impossible. 
%
% \author{%
%   D-Firstname \textsc{D-Familyname}\altaffilmark{1}
%   E-Firstname \textsc{E-Familyname}\altaffilmark{1,2}
%   and
%   F-Firstname \textsc{F-Familyname}\altaffilmark{2}}
% \altaffiltext{1}{Address of Institute}
% \email{ddddd@xxx.xxx.xx.xx}
% \email{eeeee@xxx.xxx.xx.xx}
% \altaffiltext{2}{Address of Institute}

%% `\KeyWords{}' always has to be placed before `\maketitle'.
%\KeyWords{xxxx:xxxx ......} %Do NOT move this preamble from here!

\maketitle

\begin{abstract}
With an aim of establishing how the [S/Fe] ratios behave at the
very low metallicity regime down to [Fe/H]~$\sim -3$,
we conducted a non-LTE analysis of near-IR S~{\sc i} triplet lines
(multiplet 3) at 10455--10459~$\rm\AA$ for a dozen of very metal-poor 
stars ($-3.2 \ltsim$~[Fe/H]~$\ltsim -1.9$) based on the new observational
data obtained with IRCS+AO188 of the Subaru Telescope.
It turned out that the resulting [S/Fe] values are only moderately 
supersolar at [S/Fe]~$\sim$~+0.2--0.5 irrespective of the metallicity. 
While this ``flat'' tendency is consistent with the trend recently 
corroborated by Spite et al. (2011, A\&A, 528, A9) based on the 
S~{\sc i}~9212/9228/9237 lines (multiplet 1), it disaffirms the 
possibility of conspicuously large [S/Fe] (up to $\sim +0.8$) at 
[Fe/H]~$\sim -3$ that we once suggested in our first report on the 
S abundances of disk/halo stars using S~{\sc i} 10455--10459 lines 
(Takeda \& Takada-Hidai 2011, PASJ, 63, S537). Given these new
observational facts, we withdraw our previous argument, 
since we consider that [S/Fe]'s of some most metal-poor 
objects were overestimated in that paper; the likely cause 
for this failure is also discussed.
\end{abstract}

%\section{}
%
%\noindent IMPORTANT NOTICE\\
%1. ``\verb|\draft|'' creates single column and double spaces format.\\
%2. If you comment out ``\verb|\draft|'', the output will be double column
%   and single space.\\
%3. For cross-references, the use of ``\verb|\label|, \verb|\ref|, \verb|\cite|'%' 
%   and the thebibliography environment is strongly recommended. \\
%4. Do NOT use ``\verb|\def|, \verb|\renewcommand|''.\\
%5. Do NOT redifine commands provided by PASJ00.cls.\\

%\newpage

%Sect. 1
\section{Introduction}

The galactic evolution of sulfur (one of the $\alpha$-group elements)
has been a matter of controversy, since different [S/Fe] behaviors 
with a decrease of [Fe/H] were suggested in the regime of metal-poor 
halo stars depending on the lines used; i.e., 
ever-increasing [S/Fe] even up to $\sim +0.8$ (S~{\sc i} 8693--4 
lines of multiplet 6) or nearly constant [S/Fe] at a mildly 
supersolar value around $\sim$~+0.3 (S~{\sc i} 9212/9228/9237 
lines of multiplet 1). See, e.g., Takeda et al. (2005) and the 
references therein for more details.

Given this situation, we recently carried out a systematic study 
on the [S/Fe] ratios of 33 disk/halo stars over a wide range of 
metallicity ($-3.7 \ltsim$~[Fe/H]~$\ltsim +0.3$) while newly 
exploiting the S~{\sc i} triplet lines at 10455--10459~$\rm\AA$ 
(multiplet 3) based on the near-IR spectra obtained with
Subaru IRCS+AO188, which had barely been used before 
(Takeda \& Takada-Hidai 2011; hereinafter referred to as Paper I). 
Rather unexpectedly, while the the local plateau of [S/Fe]~$\sim$~+0.2--0.4 
(flat trend) was confirmed at $-2.5 \ltsim$~[Fe/H]~$\ltsim -1.5$
in consistent with the tendency already established from the S~{\sc i}
9212/9228/9237 lines (e.g., Nissen et al. 2007), we found a considerably 
large [S/Fe] ratio amounting to $\sim$~+0.7--0.8~dex at very low 
metallicity ([Fe/H]~$\sim -3$), which apparently makes a puzzling 
discontinuity in the narrow interval of $-3 \ltsim$~[Fe/H]~$\ltsim -2.5$.
If this trend is real, the chemical evolution of sulfur would have to
be considered differently from other $\alpha$ elements generally 
showing a plateau at a mildly supersolar [$\alpha$/Fe] over the 
halo metallicity range. 

Soon after we published Paper I, however, Spite et al. (2011)
reported new results of their extensive study on the [S/Fe] 
of extremely metal-poor stars, which are markedly against 
our conclusion (i.e., considerably high [S/Fe] at [Fe/H] $\sim -3$). 
That is, using the S~{\sc i} 9212/9228/9237 lines based on the
VLT/UVES data, they showed that [S/Fe] ratio is almost constant at 
mildly supersolar values of $\sim$~0.2--0.5 over the very metal-poor 
regime of $-3.5 \ltsim$~[Fe/H]~$\ltsim -2.5$.\footnote{
J\"{o}nsson et al. (2011) also recently investigated the sulfur 
abundances of halo giants of $-2.5 \ltsim$~[Fe/H]~$\ltsim -1.5$ 
with the S~{\sc i} 10455--10459 triplet (which we used in Paper I) 
and the forbidden [S~{\sc i}] line at 10821~$\rm\AA$ based on 
the VLT/CRIRES data. They then concluded that [S/Fe] favors a flat trend
at $\sim$ +0.4 and a high [S/Fe] (i.e., rise or scatter at low [Fe/H]) 
is rather unlikely. However, since the [Fe/H] range they studied is 
limited to between $\sim -2.5$ and $\sim -1.5$ (where we also derived
a near-plateau [S/Fe]), their finding does not contradict the conclusion
of Paper I.}
How should we interpret this discordance? Do multiplet 1 lines
($\sim 0.92$~$\mu$m) and multiplet 3 lines ($\sim 1.05$~$\mu$m) 
yield different S abundances at the extremely low-metallicity regime?

Yet, we need to ascertain in the first place that the trend of high 
[S/Fe] at [Fe/H]~$\sim -3$ we obtained in Paper I from the 
S~{\sc i} 10455--10459 lines universally exists for extremely 
metal-poor stars in general, since that conclusion was extracted 
from only a few objects based on the spectra of not-so-sufficient 
quality. We thus decided to reinvestigate the [S/Fe] behavior
for a larger sample (ca. a dozen objects) specifically confined to 
{\it very} metal poor stars ($-3.2 \ltsim$~[Fe/H]~$\ltsim -1.9$) 
based on our new observations lately conducted again with Subaru 
IRCS+AO188. The purpose of this article is to report the outcome 
of this new analysis.

%Sect. 2 (table 1)
\section{Observational Data}

The near-IR spectroscopic observations were conducted on 2011 August 17 
and 18 (UT) by using IRCS+AO188 of the Subaru Telescope 
for the selected 13 very metal-poor stars, among which 4 are 
turn-off dwarfs and 9 are evolved giants. In addition, we also 
observed Vesta in order to get the sun-light reference spectrum 
with the same equipment. The list of the targets is given in table 1. 
Note that G~64-37 (which was studied in Paper I) was again included 
in the present sample.
The details of the instrument and its setting along with the
data reduction procedures (the same as in our previous observations 
in 2009 July) are described in section 2 of Paper I.

In this observing run, a special attention was paid to achieve a 
sufficiently high signal-to-noise ratio so as to detect very weak 
S~{\sc i} 10455--10459 lines of extremely metal-poor stars. 
For this purpose, especially long integrated exposure times 
($\sim$~1--4~hours) were expended to the comparatively faint 
($J \sim$~9--11~mag) lowest metallicity stars ([Fe/H]~$\sim -3$), 
such as CS~30323-048, HD~126587, G~206-34, G~64-37, HE~1523-0901, 
and BD$-$16~251. For the other brighter objects of $J \sim$~5--8 mag, 
we set the exposure times from a few minutes to $\sim 20$~min 
depending on the brightness. The S/N ratios, eventually accomplished 
in the neighborhood of the S~{\sc i} lines in our finally resulting 
$zJ$-band spectra (with the resolving power of $R\sim 20000$), 
are typically $\sim$~300--500 (cf. table 1).\footnote{Actually,
we noticed that some ripple patterns appeared in several restricted portion
of the spectrum, which may be attributed to an imperfect flat-fielding. 
While this pattern was found to fall on the region of the S~{\sc i} lines 
in several cases depending on the stellar radial velocity, we could 
successfully remove them by dividing the spectrum by that of a rapid rotator.}  

%Sect. 3 (figure 1, figure 2) 
\section{Analysis and Results}

The atmospheric parameters ($T_{\rm eff}$, $\log g$, $v_{\rm t}$, 
and [Fe/H]) necessary for constructing the model atmosphere 
for each star were taken from various published studies (cf. table 1). 
Then, as in Paper I, the S abundance was evaluated 
by way of the non-LTE spectrum-synthesis analysis while applying 
Takeda's (1995) automatic fitting procedure to the region 
of S~{\sc i} 10455--10459 lines. 

Since we modeled the observed stellar line profile ($D_{\rm obs}$) 
by the convolution of (i) the intrinsic spectrum ($D_{0}$;
where only the elemental abundance $A$ is allowed to vary 
since the model atmosphere and the microturbulence are given)
and (ii) the Gaussian macro-broadening function 
$f_{\rm M}(v)$ [$\propto \exp(-v^{2}/v_{\rm M}^{2})$]
parametrized by $v_{\rm M}$ (including the combined effects of 
the instrumental broadening,the macroturbulence, 
and the projected rotational velocity), such as 
$D_{\rm obs} = D_{\rm 0} * f_{\rm M}$, adjustable free parameters 
in accomplishing the best fit are $A$ and $v_{\rm M}$, 
both of which were actually varied in the previous analysis of Paper I.
However, for the reason mentioned in the next section, we intentionally
{\it fixed} the $v_{\rm M}$ parameter in this S~{\sc i} 10455--10459 fitting 
for five extremely metal-poor stars (CS~30323-048, HD~126587, G~206-34, 
G~64-37, HE~1523-0901, and BD$-$16~251) at the pre-determined values 
which had been established in advance from the analysis applied to 
the strong C+Si feature at $\sim 1.069$~$\mu$m [where $A$(C), $A$(Si), 
and $v_{\rm M}$ were varied to search for the best fit].
The final $v_{\rm M}$ values resulting from (or assumed as fixed in) 
the profile-fitting analysis are given in table 1.

How the theoretical spectrum for the converged solutions 
fits well with the observed spectrum is displayed in 
figure 1, and the resulting non-LTE S abundances ($A^{\rm N}$) 
along with the [S/Fe] values\footnote{
[S/Fe] $\equiv$ [S/H] $-$ [Fe/H], where 
[S/H] $\equiv A^{\rm N} - 7.20$. We used $A^{\rm N}_{\odot} = 7.20$ 
(the value derived in Paper I based on the solar flux spectrum atlas
of Kurucz et al. 1984) 
as the reference solar abundance in order to keep consistency
with our previous study, which anyhow matches the present result 
(7.21; cf. table 1) derived from the spectrum of Vesta very well.} 
are presented in table 1.

%Figure 1
\begin{figure}
  \begin{center}
    \FigureFile(80mm,80mm){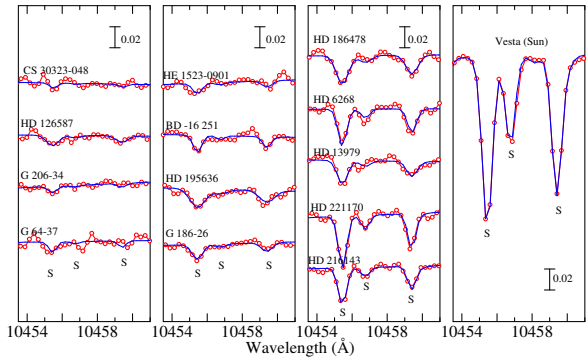}
    %%% \FigureFile(width,height){filename}
  \end{center}
\caption{Synthetic spectrum fitting for the S~{\sc i} 10455--10459
triplet lines. The best-fit theoretical spectra are shown by (blue) 
solid lines, while the observed data are plotted by (red) symbols.  
In each of the panels, spectra are arranged 
(from top to bottom, from left to right) in the ascending order 
of [Fe/H] as in table 1. A vertical offset of 0.05 is applied 
to each spectrum relative to the adjacent one, and the wavelength 
scale has been adjusted to the laboratory frame. 
}
\end{figure}

%Figure 2
\begin{figure}
  \begin{center}
    \FigureFile(80mm,80mm){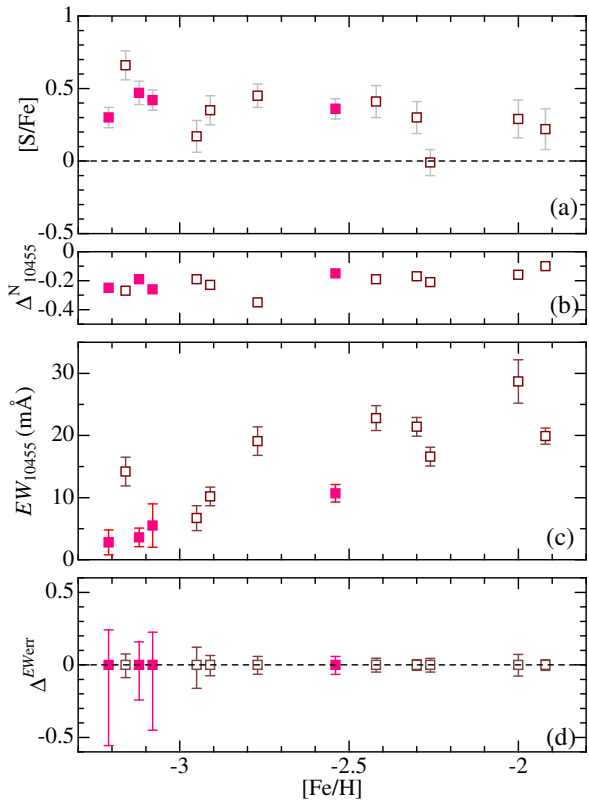}
    %%% \FigureFile(width,height){filename}
  \end{center}
\caption{Sulfur abundances and the related quantities plotted against [Fe/H]: 
(a) [S/Fe] corresponding to non-LTE sulfur abundance, where attached 
thin error bars represent the ambiguities
due to uncertainties in the atmospheric parameters ($\delta_{Tgv}$;
cf. subsection 4.2 in Paper I).
(b) $\Delta_{10455}$ (non-LTE correction for the S~{\sc i}~10455 line). 
(c) $EW_{10455}$ (equivalent width for the S~{\sc i}~10455 line)
with the S/N-dependent intrinsic random error ($\pm \delta EW$) estimated 
from Cayrel's (1988) formula.
(d) $\Delta^{EW{\rm err}}$ (abundance variation in response to $EW$ 
changes of $\pm\delta EW$).
Dwarfs ($\log g > 3$) and giants ($\log g < 3$) are indicated by filled 
(scarlet) and open (brown) squares, respectively. 
}
\end{figure}

In figure 2a are plotted the resulting [S/Fe] ratios against [Fe/H],
where the abundance uncertainties ($\delta_{Tgv}$; cf. subsection 4.2 
in Paper I) caused by ambiguities in $T_{\rm eff}$ ($\pm 100$~K), 
$\log g$ ($\pm 0.2$~dex), and $v_{\rm t}$ ($\pm 0.3$~km~s$^{-1}$) are
shown by thin error bars (though they are typically on the order of
$\sim \pm 0.1$~dex and not very significant).\footnote{
We should remark that largely different atmospheric parameters 
from those adopted here are reported in the literature, especially
for high-gravity turn-off stars. For example, Ishigaki, Chiba, and Aoki 
(2010) derived ($T_{\rm eff}$, $\log g$, [Fe/H]) of (5821, 3.50, $-3.66$) 
and (6105, 3.65, $-3.36$) for CS~30323-048 and G~64-37, respectively, 
which are considerably discrepant from Nissen et al.'s (2007) results 
of (6338, 4.32, $-3.21$) and (6432, 4.24, $-3.08$) adopted in this study; 
i.e., by $\sim$~300--500~K lower in $T_{\rm eff}$, by $\sim$~0.6--0.8~dex 
lower in $\log g$, and by $\sim$~0.3--0.5~dex in [Fe/H].
Interestingly, since the signs of the abundance corrections are opposite
(e.g., $\delta_{T-} = +0.03$~dex for $\Delta T_{\rm eff} = -100$~K,
$\delta_{g-} = -0.06$~dex for $\Delta \log g = -0.2$~dex, for G~64-37; 
cf. electronic table~E2 in Paper I), the net effect on $A$(S) itself 
is only of $\sim 0.1$~dex level and thus not so significant even in 
such extreme cases. This means, however, that [S/Fe] would be appreciably
raised because of the lowered [Fe/H], which eventually makes a serious 
discordance in comparison with the [S/Fe] results of other giants stars.}
We also derived $EW_{10455}$ (equivalent width inversely computed from 
$A^{\rm N}$) along with the corresponding non-LTE correction 
($\Delta^{\rm N}_{10455}$; negative values with typical extents of 
$\sim$~0.1--0.3~dex) for the strongest component (at 10455.45~$\rm\AA$) 
of the triplet in the same manner as described in subsection 4.2 of Paper I, 
which are given in table 1 and shown in figures 2b and 2c, respectively.
The error bars (a few m$\rm\AA$) accompanied with $EW_{10455}$ in 
figure 2c are the S/N-dependent uncertainties ($\delta EW$) estimated 
by Cayrel's (1988) formula, $\sim 1.6 (w\;\delta x)^{1/2} \epsilon$, 
where $w$ is the typical line FWHM ($\sim$~0.5--1~$\rm\AA$; assumed to 
be 0.75~$\rm\AA$), $\delta x$ is the pixel size (0.25~$\rm\AA$), and
$\epsilon$ is the photon-statistics accuracy ($\sim$~(S/N)$^{-1}$).
The abundance errors ($\Delta^{EW{\rm err}}$) in response to these 
uncertainties in $EW_{10455}$ are further displayed in 
figure 2d, where we can see that these errors become considerable 
(e.g., a few tenths dex) for extremely metal-deficient stars 
([Fe/H]~$\ltsim -3$) showing very weak lines 
($EW_{10455} \ltsim$5--6~m$\rm\AA$).

%Sect. 4 (figure 3, figure 4, figure 5)
\section{Discussion}

We are now ready to answer the question which motivated
this study: ``How do the [S/Fe] ratios of very metal-poor
stars behave around [Fe/H]~$\sim -3$: a sudden jump to a considerably high 
value of $\sim +0.8$ (as suggested in Paper I) or only a mildly 
supersolar value at $\sim +0.4$ maintaining a flat trend (as derived by 
Spite et al. 2011)?'' We can recognize in figure 2a that the [S/Fe] ratios
do not show any appreciable increase with a decrease in the metallicity,
which are almost constant over $-3.2 \ltsim$~[Fe/H]~$\ltsim -1.9$
with the mean of $\langle$[S/Fe]$\rangle = +0.34$ ($\sigma = 0.16$)
irrespective of dwarfs and giants. These [S/Fe] values of very 
metal-deficient stars resulting from this study are combined with 
those of halo/disk stars derived in Paper I on the [S/Fe] vs. [Fe/H] plot
displayed in figure 3, from which we can state that [S/Fe] ratios
are almost ``flat'' around $\sim +0.3$ over the wide metallicity range
between [Fe/H]~$\sim -1$ and $\sim -3$. Thus, we have to admit that 
Spite et al.'s (2011) argument actually represented the truth, which 
(in turn) means that our previous conclusion of Paper I was not correct, 
as far as the run of [S/Fe] around the regime of [Fe/H]~$\sim -3$ is concerned.
From an impartial point of view, however, this is a gratifying consequence
in a sense, because the different abundance indicators 
(S~{\sc i} 9212/9228/9237 and S~{\sc i} 10455--10459) eventually 
turned out to yield consistent results with each other.

%Figure 3
\begin{figure}
  \begin{center}
    \FigureFile(80mm,80mm){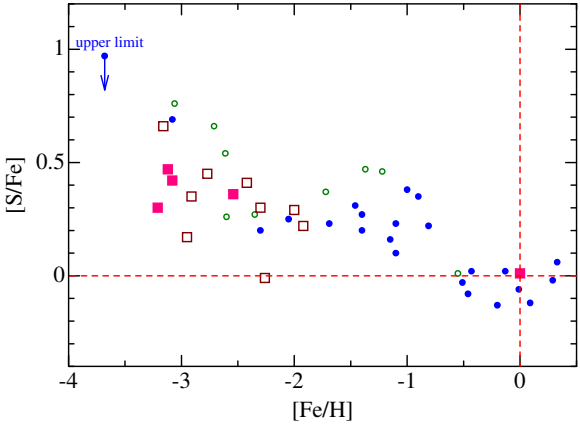}
    %%% \FigureFile(width,height){filename}
  \end{center}
\caption{
[S/Fe] vs. [Fe/H] relation based on the results in obtained this study 
(larger squares) combined with those derived in Paper I (smaller 
circles). As in figure 2, open and filled symbols are for giants 
($\log g < 3$) and for dwarfs ($\log g > 3$), respectively.
}
\end{figure}

Then, how should we interpret the prominently high [S/Fe] results
once obtained for all three most metal-poor stars around
[Fe/H]~$\sim -3$ among the sample in Paper I; i.e., 
([S/Fe], [Fe/H]) = (+0.69, $-3.08$) [G~64-37],
(+0.76, $-3.06$) [BD$-$18~5550], and (+0.66, $-2.71$) [HD~115444]?
With an intention to confirm whether or not these results are really 
reliable, we reexamined the process of the previous analysis for 
these stars.

Regarding BD$-$18~5550 and HD~115444, we noticed that the solutions
of $v_{\rm M}$ (macrobroadening parameter; cf. section 3) resulting 
as by-products of spectrum fitting are unusually high (18.6 and
19.4~km~s$^{-1}$, respectively) compared to other stars where
values around $\sim 10$~km~s$^{-1}$ are in common.\footnote{
We can roughly make an order-of-magnitude estimate of $v_{\rm M}$ 
including three kinds of macrobroadening [instrumental profile 
broadening ($v_{\rm ip}$), macroturbulence broadening ($v_{\rm mt}$),
and rotational broadening ($v_{\rm rt}$)] based on a simple 
modeling, where all three broadening functions are assumed to be 
Gaussian and each of these parameters are the corresponding $e$-folding 
width, between which the following relation holds:
$v_{\rm M}^{2} = v_{\rm ip}^{2} + v_{\rm mt}^{2} + v_{\rm rt}^{2}$.
Under this approximation, $v_{\rm ip}$ is evaluated as  
$v_{\rm ip} ={\rm FWHM}/(2\sqrt{\ln 2}) \sim 9$~km~s$^{-1}$ 
(${\rm FWHM} \sim 15$~km~s$^{-1}$ for $R \simeq 20000$).
Then, we may expect that $v_{\rm mt}$ would be as small as 
$\sim 2$~km~s$^{-1}$ according to the relation 
$v_{\rm mt} \simeq 0.4 \zeta_{\rm RT}$ (cf. footnote 12 of 
Takeda et al. 2008) since the typical $\zeta_{\rm RT}$
(radial-tangential macroturbulence) is 
$\sim 5$~km~s$^{-1}$ for early G dwarfs and early-K giants 
(cf. figure 17.10 in Gray 2005). Finally, we may regard that $v_{\rm rt}$ 
($\simeq 0.94 v_{\rm e}\sin i$; cf. footnote 12 of 
Takeda et al. 2008) is of minor importance (e.g., presumably no larger 
than $\sim$~2--3~km~s$^{-1}$ in most cases), since the stellar 
rotation must have been spun down in these old halo stars.
Accordingly, $v_{\rm M}$ is reasonably expected to be around 
$\sim 10$~km~s$^{-1}$, because it is primarily determined by the 
contribution from $v_{\rm ip}$.}
However, additional spectrum fitting analyses to the C+Si feature at 
$\lambda \sim 1.069$~$\mu$m carried out for these stars yielded quite 
reasonable $v_{\rm M}$ values of 9.4~km~s$^{-1}$ (BD$-$18~5550) and 
10.2~km~s$^{-1}$ (HD~115444) (cf. figures 4a and 4c). Since almost 
the same $v_{\rm M}$ should (in principle) result from different lines, 
we can not help considering that the $v_{\rm M}$ values for these 
two stars obtained from the S~{\sc i} 10455--10459 fitting in 
Paper I were inadequately overestimated, which we suspect may 
presumably related to the profile-fitting technique we adopted. 
That is, the automatic solution-search algorithm (Takeda 1995), 
which simultaneously varies several parameters to accomplish the best 
fit, is very efficient in case where the stellar line profiles are 
well defined. However, we should be careful when it is applied
to the very weak-line case where noises are comparable to the signal
of stellar lines, since it may yield physically meaningless solutions 
where profiles are appreciably damaged by noises. 
In such cases, $v_{\rm M}$ had better be fixed at a (more reliable) value 
derived from other stronger lines (such as the C+Si feature), instead of
varying both $A$(S) and $v_{\rm M}$, which is the approach we adopted
for the five most metal-poor stars in this study (cf. section 3).\footnote{
For the other stars, both $A$(S) and $v_{\rm M}$ were varied to
determine as in Paper I. However, we checked for each star that the 
resulting $v_{\rm M}$(S) (cf. table 1) is not in serious disagreement 
with $v_{\rm M}$ (C+Si). Note that HD~195636 is a rather unusual star, 
since its $v_{\rm M}$ (17.0~km~s$^{-1}$) is appreciably large
(an alternative C+Si fitting also yielded a similar result), which might 
have a comparatively high $v_{\rm e}\sin i$ for stars of this class.}
Accordingly, we redetermined the S abundances from the S~{\sc i} 
10455-10459 lines while {\it fixing} the $v_{\rm M}$ values at the values 
derived from the C+Si feature, and obtained $A$(S) values (4.67 and 
4.83), which are appreciably lower than the results in Paper I (4.90 and 
5.15) by 0.23~dex and 0.32~dex for BD$-$18~5550 and HD~115444,
respectively (cf. figures 4b and 4d). We thus consider that the [S/Fe] 
values of these two stars reported in Paper I should be revised downward 
by these amounts, which eventually makes +0.53 (BD$-$18~5550) and +0.34 
(HD~115444); i.e., being almost within the [S/Fe] range concluded 
in this study.

%Figure 4
\begin{figure}
  \begin{center}
    \FigureFile(80mm,80mm){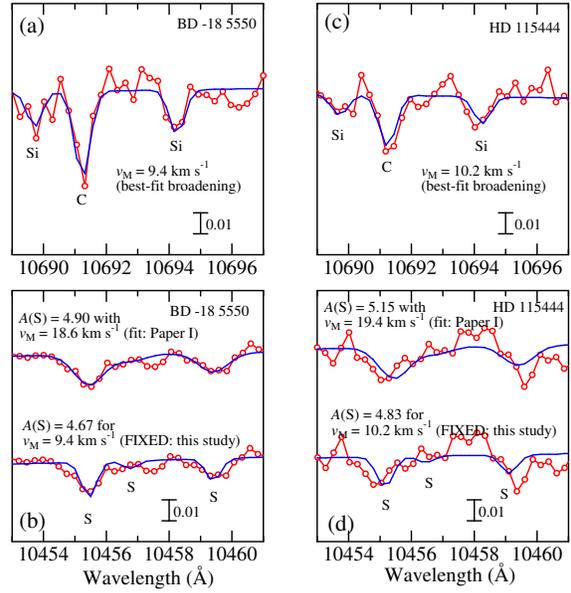}
    %%% \FigureFile(width,height){filename}
  \end{center}
\caption{
Reanalysis of two stars (BD$-$18~5550 and HD~115444 in the left 
and right panels, respectively) which showed particularly high 
[S/Fe] values in Paper I, in order to demonstrate how the choice 
of $v_{\rm M}$ (macrobroadening parameter) influences 
the resulting sulfur abundance derived from spectrum fitting.
Upper panels (a) and (c) $\cdots$ Derivation 
of best-fit $v_{\rm M}$ from the C+Si feature at $\sim 1.069$~$\mu$m. 
Lower panels (b) and (d) $\cdots$ 
Determination of $A$(S) with two kinds of $v_{\rm M}$ treatments, 
where $v_{\rm M}$ is varied as an adjustable parameter (upper
spectrum; method adopted in Paper I) or $v_{\rm M}$ is fixed at the 
value derived from the C+Si feature (lower spectrum).
}
\end{figure}

The star G~64-37 was actually an object of special attention, since
the S abundance (based on S~{\sc i} 10455--10459) derived in Paper I 
turned out appreciably higher (by $\sim$~0.6~dex) than Nissen et al.'s 
(2007) result (based on S~{\sc i} 9212/9228/9237) only for this 
extremely metal-poor star, despite that a good agreement is seen 
for other comparatively more metal-rich stars; this fact motivated us
to include this star again in our target list.
Unlike the case of BD$-$18~5550 and HD~115444, a serious mismatch
of $v_{\rm M}$ was not found in the analysis of Paper I.
Instead, we consider that the essential problem was the comparatively 
poor S/N ratio ($\sim 100$) of the spectrum used in Paper I, which is 
evidently insufficient for a reliable abundance determination from 
very weak lines with depression on the order of $\sim 1$\%.
The observed and the fitted spectrum in Paper I and those in this 
study are compared in figure 5. We can see from this figure
that the previous analysis was severely influenced by the large noise,
while the situation has been improved in present case (S/N~$\sim 200$), 
resulting in an appreciable difference of solution between the two cases. 
We naturally consider the present result of $A$(S)$=4.54$ ([S/Fe] = +0.42) 
is more credible than that in Paper I, requiring a downward revision 
of the previous [S/Fe] value (+0.69) by $-0.27$~dex, by which the 
disagreement with Nissen et al. (2007) has been reasonably mitigated 
by this amount.

%Figure 5
\begin{figure}
  \begin{center}
    \FigureFile(60mm,80mm){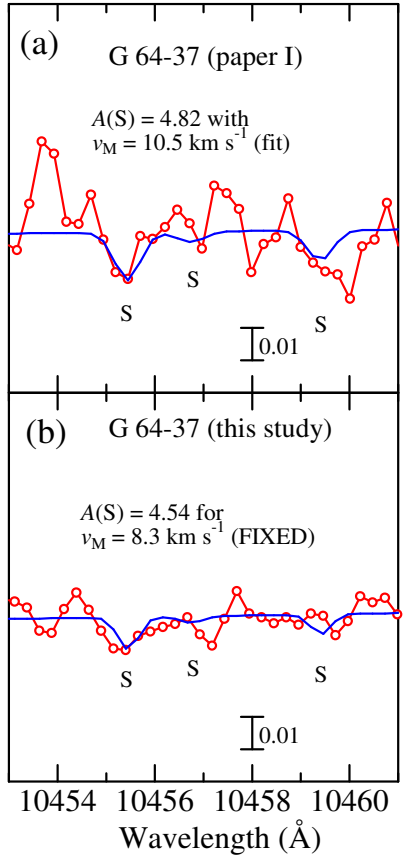}
    %%% \FigureFile(width,height){filename}
  \end{center}
\caption{
Comparison of the spectra and the $A$(S) determination procedures
for G~64-37, for which two independent analyses were performed
in Paper I (based on the 2009 data) as well as in this study
(based on the 2011 data).
}
\end{figure}

%Sect. 5
\section{Conclusion}

This study was motivated by the recent extensive work done by 
Spite et al. (2011) who reported that [S/Fe] ratios of very 
metal-poor stars determined from the S~{\sc i} 9212/9228/9237 
lines show a nearly flat behavior at a mildly supersolar 
level of $\sim$~+0.2--0.5 over a wide metallicity range of 
$-3.5 \ltsim$~[Fe/H]~$\ltsim -2$, which markedly contradict the 
conclusion of our previous study based on S~{\sc i} 10455-10459 lines 
(Paper I) that such a flat tendency of [S/Fe] (persisting down to 
[Fe/H]~$\sim -2.5$) is followed by a sudden jump up to +0.7--0.8
around [Fe/H]~$\sim -3$.

With an intention to resolve the cause of this discrepancy, 
we rechallenged the task of clarifying the [S/Fe] ratios at the
extremely low metallicity regime down to [Fe/H]~$\sim -3$ by using 
the same triplet lines as used in Paper I, based on the new observational
data for an extended sample of 13 very metal-poor stars observed
with IRCS+AO188 of the Subaru Telescope. 

In almost the same manner as in Paper I, we conducted a non-LTE 
spectrum fitting analysis of S~{\sc i} 10455--10459 triplet,
and found that the resulting [S/Fe] values were moderately 
supersolar uniformly scattering around $\sim$~+0.3--0.4 
[with the mean abundance of $\langle$[S/Fe]$\rangle$~=~+0.34 
($\sigma = 0.16$)] over $-3.2 \ltsim$~[Fe/H]~$\ltsim -1.9$ 
without any systematic [Fe/H]-dependence,
which confirmed the consequence corroborated by Spite et al. (2011) 
based on the S~{\sc i} 9212/9228/9237 lines.

Given these new observational facts, we reexamined the process of 
our previous analysis for the three extremely metal-poor stars 
(G~64-37, BD$-$18~5550, and HD~115444), for which we derived prominently
high [S/Fe] values ($\sim$~+0.7--0.8) that eventually lead to 
the conclusion of Paper I. Regarding G64-37, our reanalysis using
a new spectrum of higher quality yielded a result lower than the
previous value by $\sim 0.3$~dex. For BD$-$18~5550 and HD~115444,
we noticed that our automatic profile-fitting method (which varies 
both the abundance and the broadening parameter to find the 
best fit) resulted in unreasonably large solutions of the 
broadening width, because of the considerable weakness of
the line profile severely damaged by noises. When the broadening
parameter was fixed at the more reasonable values determined from 
stronger lines, we found that the revised solution of the S abundance 
is lowered by $\sim$~0.3--0.4~dex for both of these stars.
Consequently, it is likely that we had overestimated the [S/Fe] 
values of these three stars in Paper I by $\sim$~0.3--0.4~dex.

Accordingly, we now consider that the flat trend of [S/Fe] 
(without any systematic rise with a decrease of metallicity) represents 
the truth, at least with regard to the overall [S/Fe] behavior of 
very metal-poor stars in general, which means the withdrawal of 
our previous argument that [S/Fe] experiences a sudden jump 
up to conspicuously large [S/Fe] of $\sim +0.8$ as the metallicity is
lowered down to the [Fe/H]~$\sim -3$ regime.

We finally remark, however, that the existence of some stars deviating 
from the main trend with appreciably higher/lower [S/Fe] amounting to 
 [S/Fe]~$\sim$~+0.7/0.0 as a result of the natural diversity is not 
necessarily be excluded (e.g., we derived [S/Fe]~=~+0.66 for 
HD~126587 and [S/Fe]~=~$-0.01$ for HD~13979). 
Actually, since our 13 stars (at [Fe/H]~$\ltsim -2$) yielded 
$\langle$[S/Fe]$\rangle = +0.34$ with $\sigma$ (standard deviation) 
of 0.16, we may expect the probability of finding $\ge 1\sigma$  
deviation ([S/Fe]~$\ltsim +0.2$ or [S/Fe]~$\gtsim +0.5$) and 
$\ge 2\sigma$ deviation ([S/Fe]~$\ltsim 0.0$ or [S/Fe]~$\gtsim +0.7$)   
to be $\sim 30$\% and $\sim 5$\%, respectively.\footnote{
This estimation naturally depends on the extent of $\sigma$.
We note that the $\sigma$([S/Fe]) of [Fe/H]~$\le -2$ stars
derived by other investigators are somewhat smaller than our result;
e.g., $\sigma = 0.07$ (Nissen et al. 2007), $\sigma = 0.12$ 
(Spite et al. 2011), $\sigma = 0.11$ (J\"{o}nsson et al. 2011). 
It should thus be important to quantitatively establish not only 
the trend of [S/Fe] on the average but also the extent of its 
dispersion based on a large sample of very metal-poor stars.}
Form this point of view, the recent Koch and Caffau's (2011) result 
of [S/Fe] as somewhat high as $\sim +0.5$ for a red giant in the 
very metal-poor globular cluster NGC~6397 ([Fe/H]~$\sim -2.1$) may 
be understandable without invoking a bimodal [S/Fe] trend such as 
they discussed.

\bigskip

We express our heartful thanks to T.-S. Pyo and Y. Minowa
for their kind advices and helpful support in preparing
as well as during the IRCS+AO188 observations.

One of the authors (M. T.-H.) is grateful for a financial support 
from a grant-in-aid for scientific research (C, No. 22540255) 
from the Japan Society for the Promotion of Science.

This research has made use of the SIMBAD database, operated by
CDS, Strasbourg, France.

\clearpage

\onecolumn

%Table 1
\clearpage
\setcounter{table}{0}
\begin{table}[h]
\caption{Parameters of the program stars and the results of abundance analyses.}
\scriptsize
\begin{center}
\begin{tabular}{cc@{ }c@{ }c@{ }c c rrrcccl} 
\hline\hline
Name & $T_{\rm eff}$ & $\log g$ & $v_{\rm t}$ & [Fe/H] & Ref. & $v_{\rm M}$ & $A^{\rm N}$ & 
$EW_{10455}$ & $\Delta_{10455}$ & [S/Fe] & S/N & Remark \\
 & (K) & (cm~s$^{-2}$) & (km~s$^{-1}$) & (dex) &  & (km~s$^{-1}$) & (dex) & (m$\rm\AA$) &(dex)&(dex)& & \\
\hline
CS~30323-048 &6338& 4.32& 1.5& $-$3.21 & NIS07 & (8.7) & 4.29 &  2.8& $-$0.25& +0.30& 350 & $v_{\rm M}$ fixed, larger uncertainty \\
HD~126587    &4700& 1.05& 1.7& $-$3.16 & HAN11 & (12.7) & 4.70 & 14.2& $-$0.27& +0.66& 300 & $v_{\rm M}$ fixed \\
G~206-34     &5825& 3.99& 1.5& $-$3.12 & RIC09 & (9.4) & 4.55 &  3.6& $-$0.19& +0.47& 450 & $v_{\rm M}$ fixed \\
G~64-37      &6432& 4.24& 1.5& $-$3.08 & NIS07 & (8.3) & 4.54 &  5.5& $-$0.26& +0.42& 200 & $v_{\rm M}$ fixed \\
HE~1523-0901 &4630& 1.00& 2.6& $-$2.95 & FRE07 & (10.3) & 4.42 &  6.7& $-$0.19& +0.17& 350 & $v_{\rm M}$ fixed \\
BD$-$16~251  &4825& 1.50& 1.8& $-$2.91 & AND10 & 9.4 & 4.64 & 10.2& $-$0.23& +0.35& 450 & \\
HD~195636    &5370& 2.40& 1.5& $-$2.77 & CAR03 & 17.0 & 4.88 & 19.1& $-$0.35& +0.45& 300 & large $v_{\rm M}$ (cf. footnote 5)\\
G~186-26     &6417& 4.42& 1.5& $-$2.54 & NIS07 & 12.0 & 5.02 & 10.7& $-$0.15& +0.36& 500 & \\
HD~186478    &4730& 1.50& 1.8& $-$2.42 & HAN11 & 13.3 & 5.19 & 22.8& $-$0.19& +0.41& 350 & \\
HD~6268      &4735& 1.61& 2.1& $-$2.30 & SAI09 & 9.5 & 5.20 & 21.4& $-$0.17& +0.30& 450 & \\
HD~13979     &5075& 1.90& 1.3& $-$2.26 & BUR00 & 11.8 & 4.93 & 16.6& $-$0.21& $-$0.01& 450 & \\
HD~221170    &4560& 1.37& 1.6& $-$2.00 & SAI09 & 9.2 & 5.49 & 28.7& $-$0.16& +0.29& 200 & \\
HD~216143    &4525& 1.77& 1.9& $-$1.92 & SAI09 & 9.4 & 5.50 & 19.9& $-$0.10& +0.22& 550 & \\
\hline
Vesta (Sun)   &5780& 4.44& 1.0&    0.00 & $\cdots$ & 10.7 & 7.21 &123.5& $-$0.09& +0.01& 600 & \\
\hline
\end{tabular}
\end{center}
In columns 1 through 6 are given the star designation, 
effective temperature, logarithmic surface gravity, 
microturbulent velocity dispersion, Fe abundance relative to
the Sun, and key for the reference of atmospheric parameters:
BUR00 $\cdots$ Burris et al. (2000), 
CAR03 $\cdots$ Carney et al. (2003),
FRE07 $\cdots$ Frebel et al. (2007), 
HAN11 $\cdots$ Hansen and Primas (2011),
NIS07 $\cdots$ Nissen et al. (2007),
RIC09 $\cdots$ Rich and Boesgaard (2009),
SAI09 $\cdots$ Saito et al. (2009). 
Columns 7--11 present the results of the abundance analysis
based on the S~{\sc i} 10455--10459 profile-fit: $v_{\rm M}$ is the 
best-fit macrobroadening parameter (while those in parentheses are 
the assumed or fixed values, which were separately derived 
from the C+Si 1.069~$\mu$m feature fitting),
$A^{\rm N}$ is the non-LTE logarithmic abundance of S 
(in the usual normalization of H = 12.00) derived from spectrum-synthesis 
fitting, $EW_{10455}$ is the equivalent width (in m$\rm\AA$) for the 
S~{\sc i} 10455 line inversely computed from $A^{\rm N}$, $\Delta_{10455}$ 
is the non-LTE correction ($\equiv A^{\rm N} - A^{\rm L}_{10455}$) for the 
S~{\sc i} 10455 line, and [S/Fe] ($\equiv A^{\rm N}$ $-$ 7.20 $-$ [Fe/H]) 
is the S-to-Fe logarithmic abundance ratio relative to the Sun. 
The S/N ratio of the spectrum estimated at the position of the S~{\sc i} 
lines is given in column 12.
The objects are arranged in the ascending order of [Fe/H]. 
\end{table}

\end{document}